\documentstyle[aps,epsf]{revtex}
\begin{document}
\large
\title{\LARGE \bf Binding and Off-Mass-Shell Effects \\
in Deep Inelastic Scattering on Deuteron.}
\author{\Large V.V.~Burov and A.V.~Molochkov\\[.5cm]}
\address{\large BLTP, Joint Institute for Nuclear Research, 141980 Dubna,
Russia\\[.5cm]}
\maketitle
\begin{abstract}
The role of relativistic off-mass-shell kinematics
in the deep inelastic scattering on deuteron is analyzed.
It is shown that the relativistic impulse approximation
reproduces effects from the binding, Fermi motion, and two-nucleon
contribution. The nonrelativistic limit of the deuteron structure
function is in agreement with the nonrelativistic calculations.\\[1cm]
\end{abstract}

The discovering of the EMC-effect initiated active studying of
nuclear effects in the deep inelastic scattering (DIS) on the deuteron
from the point of view of QCD and nuclear physics.
These investigations allow one to rely that nuclear
properties can be expressed in terms of the QCD parameters~\cite{Close}.
On the other hand, successful description of deep inelastic
scattering on deuteron can give us information about the neutron
structure function~\cite{Grossn,param}.
All these tasks demand consistent calculations of
nuclear effects in DIS.

Due to recent success in the description of the deuteron low energy
characteristics
in the relativistic formalism \cite{tjon,bond}
one hopes to construct in the framework of this formalism a consistent
picture of different processes on the deuteron.
In the present paper, we analyze the deep inelastic scattering on the
deuteron in the framework of the approach based on
the Bethe-Salpeter formalism.

\section{Basic approximations}
Most of the relativistic field theory approaches to DIS on nuclei
are based on the following basic approximations~\cite{Gross,KUH}:

\begin{itemize}
\item  Ladder approximation for the deuteron vertex function;
\item  Disregard of the interference terms in the square of
       the scattering amplitude;
\item  The same representation for the nucleon amplitude via scalar structure
functions for both a bound and a free nucleon.
\end{itemize}

 The first approximation makes it possible to solve the Bethe-Salpeter
 equation in the mesonic nucleon theory \cite{tjon,KUH}.
In this approximation, the interaction between
nucleons is represented as an infinite ladder of one-mesonic exchanges.
It has been shown that this approximation allows one to describe the basic
properties of deuteron such as its mass, binding energy and
magnetic and quadruple momenta simultaneously.\cite{tjon,bond}
 This approximation
works well when the relative energy of constituents is small.
Presupposing this in the deuteron we keep this approximation.

 The second approximation permits one to consider the DIS amplitude as
an incoherent sum
of squared terms corresponding to the scattering on individual constituents.
This approximation possibly
works well in the Bjorken limit. Calculation at
the finite $Q^2$  and large and low $x$ requires a detailed analysis of
the interference terms.
A simple analysis of the Compton amplitude shows that most of the interference
terms are suppressed as an additional power of $1 \over Q^2$~\cite{deut}.
 Nevertheless, this approximation needs a more explicit investigation.

 The third approximation was analyzed in several papers which considered
additional structure functions in the representation for the hadron tensor.
The analysis motivated by the Sullivan model \cite{suliv,kapoff,melnt}
has shown that these additional structure functions can be
neglected, and off-mass-shell nucleon amplitudes can be replaced by
the on-mass-shell one.
\begin{eqnarray}
W_{\mu \nu}(P,q)=\left( -g_{\mu\nu} + \frac{q_\mu q_\nu}{q^2}\right)F_1(x) +
\frac{1}{P\cdot q}\left(P_\mu - \frac{P\cdot q}{q^2}q_\mu\right)
\left(P_\nu - \frac{P\cdot q}{q^2}q_\nu\right)F_2(x). \label{lorentz}
\end{eqnarray}
 In our opinion this approximation is not so clear and even if we could
represent the hadron tensor through two structure functions,
both of them are
not indentical to the on-mass-shell one.
In the paper \cite{deut}, we have shown that this approximation leads to
the representation for the deuteron hadron tensor via
convolution of the nucleon structure function and distribution function,
satisfying the baryon sum rule:
$$\int \frac{d^4k}{(2\pi)^4}f(M_D,k)=2$$
 It follows from this relation that
if we neglect the off-mass-shell effects and identify off-shell and on-shell
nucleon structure functions, the ratio of the deuteron
and isoscalar nucleon structure function becomes equal to $1$
when $x\rightarrow 0$.
However, application of this approximation gives no EMC-like behavior of
this ratio, what means it has no nuclear effects
but the Fermi motion. An attempt to introduce the dependence
on the relative energy $k_0$
conserving on-mass-shell form of the hadron tensor and structure
functions leads to changing of the ratio at $x=0$ and thus to
contradiction with the baryon sum rule.

We show in this paper how this contradiction can be removed by the
expansion of the convolution formula near the nucleon mass-shell.
Moreover inclusion of
the off-mass-shell effects for the bound nucleon structure function
reproduces the EMC-effect for the deuteron and the two-nucleon contributions.

\section{Compton amplitude.}

Following Bethe-Salpeter based approach \cite{deut} we can write
the forward Compton amplitude on deuteron as
\begin{eqnarray}
T_{\mu \nu}^{D}(P,q)=\int \frac{d^4k}{(2\pi)^4}\overline{\Gamma}(P,k)
{\overline{G}_6}_{\mu\nu}(P,k,q)\Gamma(P,k).
\label{compt}\end{eqnarray}
Here $\Gamma(P,k)$ is the Bethe-Salpeter vertex function defined as
\begin{eqnarray}
\Gamma(P,k)={S^{(2)}}^{-1}(P,k)\int d^4x d^4X e^{ikx}e^{iPX}
\langle 0|{\rm T}(\psi (X-x/2)\psi (X+x/2))|D\rangle
\nonumber\end{eqnarray}
and satisfying
the homogeneous Bethe-Salpeter equation~\cite{BS}:
\begin{eqnarray}
\Gamma^S(P,k)=-\int \frac{d^4k^{\prime}}{(2\pi)^4} \overline{G}_4
(P,k,k^{\prime}){S^{(2)}}(P,k^\prime)
\Gamma^S(P,k^{\prime})\label{BS}.
\end{eqnarray}
The normalization condition is fixed as the
deuteron electromagnetic current
at zero momentum transfer:
\begin{equation}
\langle D| J_\mu(0) |D \rangle = 2iP_\mu.
\end{equation}
The $\overline{G}_4(P,k,k^{\prime})$ and ${\overline{G}_6}_{\mu\nu}(P,k,q)$
denote the irreducible two-nucleon Green functions. The second is two-nucleon
Green function with
the insertion of a T-product of the nucleon electromagnetic current operators.
Formally, the irreducible kernel for the Bethe-Salpeter equation  can be
defined via the exact two-nucleon Green function $G_4(P,k,k^\prime)$:
\begin{eqnarray}
G_4(P;k,k^\prime)&=&S^{(2)}(P,k)\left((2\pi)^4\delta(k-k^\prime)+
\phantom{\frac{dx}{(2\pi)^4}}\right.\label{g4}\\
&&\left. + \sum\limits_{n\ge 1}\frac{1}{n!}\int\frac{d^4k_1}{(2\pi)^4}...
\frac{d^4k_n}{(2\pi)^4}\overline{G}_4(P;k,k_1)S^{(2)}(P,k_1)...
\overline{G}_4(P;k_n,k^\prime)S^{(2)}(P,k^\prime)\right)
\nonumber\end{eqnarray}
where $S^{(2)}(P,k)$ is a direct product of the two nucleon propagators.
The irreducible two-nucleon Green function with the insertion of the T-product
of the nucleon electromagnetic current operators can be
defined as:
\begin{eqnarray}
\overline{G_6}_{\mu\nu}(q,P,k^\prime,k)=
\int \frac{d^4k_1}{(2\pi)^4} \frac{d^4k_2}{(2\pi)^4}
G^{-1}_4(P;k,k_1){G_6}_{\mu\nu}(q,P,k_1,k_2)G^{-1}_4(P;k_2,k^\prime)
\end{eqnarray}
where
\begin{eqnarray}
{G_6}_{\mu\nu}(q,P,k^\prime,k)&=& i\int d^4x d^4y d^4y^\prime d^4Y d^4Y^\prime
e^{-iky+ik^\prime y^\prime} e^{-iqx} e^{-iP(Y-Y^\prime)} \\ \nonumber
&&<0|{\rm T}({\bar \psi}(Y+\frac y2){\bar \psi}(Y-\frac y2)J_\mu (x) J_\nu (0)
\psi (Y^\prime+\frac {y^\prime}{2}) \psi (Y^\prime-\frac {y^\prime}{2}))|0>.
\end{eqnarray}
The diagrams on fig.1. schematically present different contributions
to the forward Compton scattering amplitude on the deuteron.
The term a) corresponds to the relativistic impulse approximation.
In the ladder approximation, the term f) contains only mesonic exchange
current contribution. Both of them define the
convolution approximation.
The term f) affects
the behavior of the structure function only at small  $x$.
Thus, in the relativistic convolution approximation there are no
mesonic corrections ensuring the EMC-behavior.
\vspace*{-.1cm}
\begin{figure}[h]
\epsfxsize=7.5cm
\hspace*{7.8cm}\epsfbox{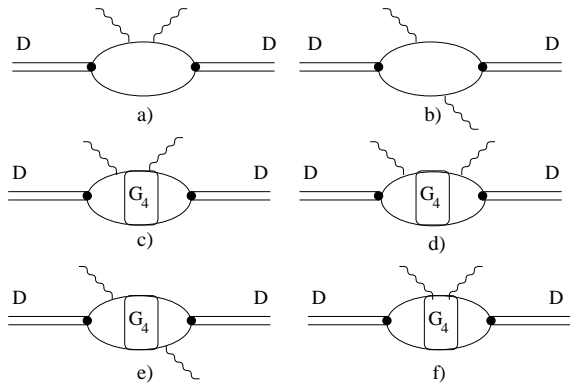}
\hspace*{-8cm}
\begin{minipage}{8.1cm}
\vspace*{3cm}\caption{Diagrams schematically present the
compton scattering on the deuteron}
 \label{fig1}
 \end{minipage}
 \end{figure}
\vspace*{-7.4cm}\noindent
\begin{minipage}{7.5cm}
\indent \phantom{aa}Analyzing terms behind the
con\-vo\-lu\-ti\-on ap\-pro\-xi\-ma\-ti\-on,
we can note that
their imaginary part contains a nucleon propagator with
high momentum. So it has an additional power of $\frac{1}{Q^2}$.
We cannot say that about the terms b) and e).
But we can assume
that these terms can be suppressed because of high nucleon relative
energy. In any case, these terms correspond to the scattering on the deuteron
as on a whole and we presuppose that they will be visible at high $x$ or
finite $Q^2$. Thus if we neglect the terms e) and b), we get only
the relativistic impulse approximation to describe the EMC-behavior
of the deuteron and nucleon structure function ratio.
\end{minipage}
\phantom{aaa}
\newpage

\section{Relativistic impulse approximation.}

Taking only the term in (\ref{compt})
corresponding to relativistic impulse approximation
we get the deuteron hadron tensor in the form:
\begin{eqnarray}
\label{imp21} W_{\mu \nu }^{D}(P,q)=\int \frac{d^4k}{(2\pi )^4}W_{\mu \nu
}^N\left(\frac P2+k,q\right)f^N(P,k) +
\int \frac{d^4k}{(2\pi )^4}W_{\mu \nu
}^{\overline{N}}\left(\frac P2+k,q\right)f^{\overline{N}}(P,k)\label{hadron}
\end{eqnarray}
where $k$ is the relative momentum
of nucleons inside the deuteron.
The distribution functions $f^N(P,k)$ have the following form:

\begin{eqnarray}
f^N(P,k)=
\frac{im^2}{2E^3}\frac{1}{\left(\frac{M_D}{2}+k_0-E\right)^2}
\left[\frac{\Phi^2_{++}(P,k)}{-k_0-(E-\frac{M_D}{2})}+
\frac{\Phi^2_{+-}(P,k)}{-k_0+(E+\frac{M_D}{2})}\right]\nonumber\\
f^{\overline N}(P,k)=
\frac{im^2}{2E^3}\frac{1}{\left(\frac{M_D}{2}+k_0+E\right)^2}
\left[\frac{\Phi^2_{--}(P,k)}{-k_0+(E+\frac{M_D}{2})}+
\frac{\Phi^2_{-+}(P,k)}{-k_0-(E-\frac{M_D}{2})}\right],\nonumber
\end{eqnarray}
The functions $\Phi$ are expressed via the BS-vertex function
$\Gamma(P,k)$:
\begin{eqnarray}
\Phi^2_{++}(M_D,k)&=&\sum_S\overline{\Gamma}^S_{\alpha\beta}(M_D,k)
\sum_s u^s_\alpha({\bf k})\overline{u}^s_\delta({\bf k})
\sum_s u^s_\beta(-{\bf k})\overline{u}^s_\gamma(-{\bf k})
\Gamma^S_{\delta\gamma}(M_D,k)\nonumber \\
\Phi^2_{+-}(M_D,k)&=&-\sum_S\overline{\Gamma}^S_{\alpha\beta}(M_D,k)
\sum_s u^s_\alpha({\bf k})\overline{u}^s_\delta({\bf k})
\sum_s v^s_\beta({\bf k})\overline{v}^s_\gamma({\bf k})
\Gamma^S_{\delta\gamma}(M_D,k)\nonumber \\
\Phi^2_{-+}(M_D,k)&=&-\sum_S\overline{\Gamma}^S_{\alpha\beta}(M_D,k)
\sum_s v^s_\alpha(-{\bf k})\overline{v}^s_\delta(-{\bf k})
\sum_s u^s_\beta(-{\bf k})\overline{u}^s_\gamma(-{\bf k})
\Gamma^S_{\delta\gamma}(M_D,k)\nonumber \\
\Phi^2_{--}(M_D,k)&=&\sum_S\overline{\Gamma}^S_{\alpha\beta}(M_D,k)
\sum_s v^s_\alpha(-{\bf k})\overline{v}^s_\delta(-{\bf k})
\sum_s v^s_\beta({\bf k})\overline{v}^s_\gamma({\bf k})
\Gamma^S_{\delta\gamma}(M_D,k)\nonumber \\
\end{eqnarray}

The elementary nucleon (antinucleon) amplitude $W_{\mu \nu}^{N(\overline{N})}
\left(\frac P2+k,q\right)$ differs from the amplitude of a physical nucleon.
It appears in the dependence on the time component of the nucleon
relative momentum,
and in the difference between the total momentum the square of
this nucleon and square of
its mass $m$.
Also, it cannot be represented via the two structure functions only.
To expand expression (\ref{imp21}) in terms of the on-mass-shell nucleon
hadron tensor, we presuppose the leading contribution of
the amplitude near the mass-shell of the nucleon interacting with the photon
and approximate expression (\ref{hadron})
by the residue at $\frac{M_D}{2}+k_0=E$ for the nucleon contribution part and
at $\frac{M_D}{2}+k_0=-E$ for the antinucleon:
\begin{eqnarray}
W_{\mu\nu}^D(M_D,q)&=&\int\frac{d^3k}{(2\pi)^3}\frac{m^2}{2E^3(M_D-2E)^2}
\left\{\phantom{\frac{a^2}{b^2}}
\Phi^2_{++}(M_D,k)W_{\mu\nu}^N({\bf k},q)+ \right. \label{hadr} \\
&&\left. (M_D-2E)\frac{\partial}{\partial k_0}\left(W_{\mu\nu}^N(k,q)
\Phi^2_{++}(M_D,k)\right)_{k_0=k^N_0}+ \right. \nonumber \\ &&\left.
\frac{(M_D-2E)^2}{M_D^2}\left[\phantom{\frac{a^2}{b^2}}
\Phi^2_{+-}(M_D,k)W_{\mu\nu}^N({\bf k},q)+
\Phi^2_{-+}(M_D,k)W_{\mu\nu}^{\overline{N}}({\bf k},q)+
\right.\right.\nonumber\\ &&\left. \left.
\hspace*{-2.1cm}M_D\frac{\partial}{\partial k_0}\left(W_{\mu\nu}^N(k,q)
\Phi^2_{+-}(M_D,k)\right)_{k_0=k^N_0}+
M_D\frac{\partial}{\partial k_0}\left(W_{\mu\nu}^{\overline{N}}(k,q)
\Phi^2_{-+}(M_D,k)\right)_{k_0=k^N_0}+\right. \right. \nonumber \\
&&\left. \left.
\hspace*{-2.2cm}\frac{M_D^2}{(M_D+2E)}
\frac{\partial}{\partial k_0}\left(W_{\mu\nu}^{\overline{N}}(k,q)
\Phi^2_{--}(M_D,k)\right)_{k_0=k^N_0}+
\frac{M_D^2}{(M_D+2E)^2}\Phi^2_{--}(M_D,k)W_{\mu\nu}^{\overline{N}}
({\bf k},q)\right]
\right\}\nonumber
\end{eqnarray}

As it follows from the energy conservation law, the construction
$M_D-2E$ is the kinematic expression of a binding potential.
Thus, we can conclude that the contribution of the antinucleon states
is suppressed as an additional power of the potential.
The contribution of the BS-vertex function derivative over the relative
energy in the relativistic impulse approximation obviously is connected
with the two-nucleon contribution.

We can observe here that expression (\ref{hadr}) contains only
a free nucleon hadron tensor and its derivatives at the nucleon mass-shell.
Thus, neglecting the derivatives of additional structure functions in the
representation
of the hadron tensor, we can get the structure function
$F_2^D$ in the Bjorken limit:
$$\lim _{Q^2\rightarrow \infty }g_{\mu \nu }W^{\mu \nu }(P,q)
=-\frac 1x F_2(x).$$
$$\frac{\partial}{\partial k_0}
\lim _{Q^2\rightarrow \infty }g_{\mu \nu }W^{\mu \nu }(P,q)
=\left[\frac{1}{x^2}F_2(x)-\frac{1}{x}\frac{d}{dx}F_2(x)\right]
\left(\frac{\partial x}{\partial k_0}\right)_{k_0=k^N_0}$$
Neglecting terms of $(M_D-2E)^2$ order we obtain following expression
for the deuteron structure function:
\begin{eqnarray}
F_2^D(x_D)=\int \frac{d^3k}{(2\pi)^3}
\frac{m^2}{4E^3(M_D-2E)^2}\left\{F_2^N(x_N)
\left(1-\frac{E+k_3}{M_D}\right)\Phi^2(M_D,k)
- \right. \label{f2}
\end{eqnarray}
\vspace*{-.5cm}
\begin{eqnarray}
\left. \frac{M_D-2E}{M_D}
x_N\frac{dF_2^N(x_N)}{dx_N}\Phi^2(M_D,k)
+ F_2^N(x_N)\frac{E-k_3}{M_D}(M_D-2E)
\frac{\partial}{\partial k_0}\Phi^2(M_D,k)\right\}_{k_0=E-\frac{M_D}{2}}
\nonumber\end{eqnarray}

\section{Nonrelativistic Limit}
To compare our result with the nonrelativistic calculations,
we expand $E$ in powers of $\frac{{\bf p}^2}{m^2}$ in
(\ref{f2}) and discard the pure relativistic contributions connected with
the two-nucleon effects.
This gives us the following expression for the structure function $F_2^D$:
\begin{eqnarray}
F_2^D(x_D)=\int \frac{d^3k}{(2\pi)^3}
\left\{F_2^N(x_N)
\left(1-\frac{k_3}{m}\right)
\Psi^2({\bf k})
-
\right. \phantom{aaaaaaaaaaaaaaa}\label{f2nr}
\end{eqnarray}
\vspace*{-.5cm}
\begin{eqnarray}
\phantom{aaaaaaaaaaaaaaaaaaaaaaaaaaaaaa}\left.
\frac{-T+\epsilon}{m}
x_N\frac{dF_2^N(x_N)}{dx_N}\Psi^2({\bf k})\right\}
\nonumber\end{eqnarray}
$T=2E-2m$ is the kinetic energy of nucleons, and $\epsilon=M-2m$ is the binding
energy.

Here, we introduce an analog of the nonrelativistic wave function
$\Psi^2({\bf k})$:
$$
\Psi^2({\bf k})=\frac{m^2}{4E^2M_D(M_D-2E)^2}
\left\{\Phi^2(M_D,k)\right\}_{k_0=E-\frac{M_D}{2}}
$$
with the usual normalization condition
$$\int \frac{d^3k}{(2\pi)^3}\Psi^2({\bf k})=1.$$
Comparing (\ref{f2nr}) with the nonrelativistic calculations~\cite{param},
 we can conclude
that we have got an analog of the nonrelativistic
impulse approximation with interaction corrections.
Distinction from calculation \cite{param} consists only in convolution
with the distribution function in the second term containing the derivative
of the nucleon structure function.
\newpage

\begin{figure}[t]
\epsfxsize 10cm
\vspace*{-6cm}
\epsfbox{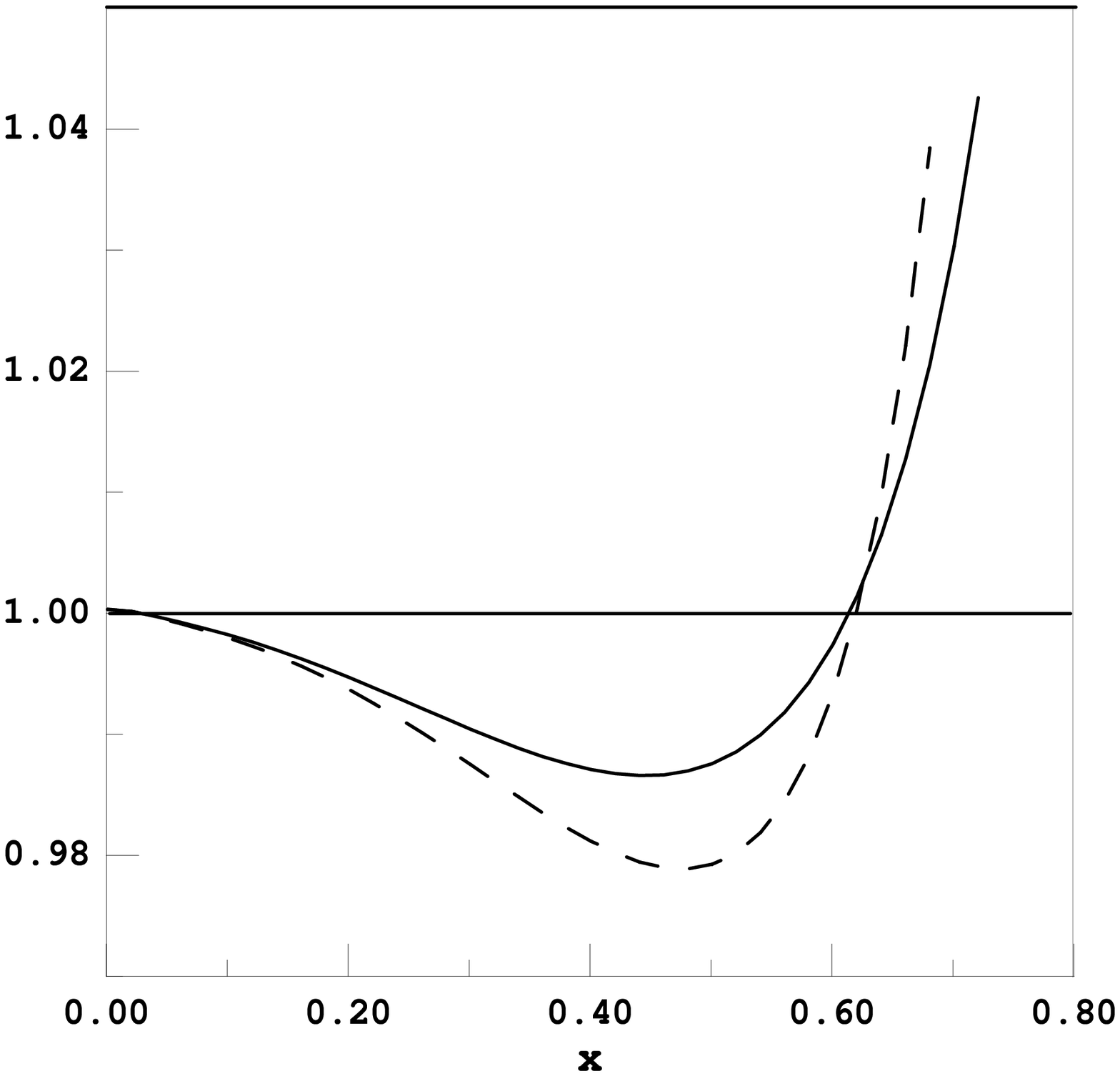}
\vspace*{-8cm}
\hspace*{10cm}\begin{minipage}{5.5cm}
Figure 2. Ratio of deuteron and isoscalar nucleon structure functions.
Continuous curve is the calculation by (\ref{f2nr}), dashed curve is
the relativistic calculation from \cite{kopen}
with mesonic degrees of freedom.
\end{minipage}
\end{figure}
\phantom{aa}
\vspace*{6cm}

\section{Numerical results}

Numerical calculations of (\ref{f2nr}) are presented on fig.2. by the
continuous curve. The calculation was performed with the Bethe-Salpeter
vertex function calculated in the approach with the separable form
of interaction~\cite{RuppTjon}. The parametrization for
the isoscalar nucleon structure function is taken from~\cite{param}.

At small $x\simeq 0$ the ratio of the deuteron and nucleon structure functions
equals $1$. It is connected with the neglect of the two-nucleon contribution
expressed by the derivative with respect to relative energy
of the Bethe-Salpeter vertex function, and the relativistic mesonic exchange
currents which affect the ratio at small $x$.

At intermediate $x$ ($0.1\le x \le 0.6$) the deuteron structure
function is suppressed
in comparison with the nucleon one. This suppression is a corollary of the
off-mass-shell
behavior of the nucleon structure function, which is ensured by the derivative
of the nucleon structure function in expression (\ref{f2}).
This result is in qualitative agreement with calculation including
the binding by the mesonic corrections~\cite{kopen}, which
is presented by the dashed curve on figure 2.

The relativistic Fermi motion provides a rapid growth of the ratio
at large $x\ge 0.6$. The ratio obtained from (\ref{f2}) has a softer
growth in this region of $x$ and smaller EMC-effect at
intermediate $x$. It is connected with the distribution
over the whole region of $x$
of the contribution of the structure function derivative.

\section{Conclusion.}

In this paper we have analyzed the connexion between
the off-mass-shell kinematics of nucleons inside the deuteron and
the binding effects in the deep inelastic scattering on the deuteron.
It is shown that assuming nucleons near the mass-shell we can express the
deuteron structure function $F_2^D$ in terms of a free nucleon
structure function and its derivatives.
The contribution of $P$, $S^{--}$, $D^{--}$ waves proves to be
suppressed as a second power of nucleon-nucleon potential in
comparison  with the $S^{++}$, $D^{++}$ wave contribution.

Comparison of the obtained results with the previous nonrelativistic
calculations shows that the influence of mesonic degrees of freedom in
the nonrelativistic models,
leading to binding effects, can be reproduced in the relativistic impulse
approximation in the framework of the Bethe-Salpeter formalism. So we can
conclude that these corrections are to some extent nonrelativistic
parametrization of the off-mass-shell effects.

As it has been assumed in this paper, relativistic mesonic corrections do not
influence the EMC behavior of the ratio of the deuteron
and nucleon structure functions
ratio. Possibly, they can significantly
influence at large and low $x$.
Further investigations of terms beyond the convolution approximation
can help one answer this question and
give a description of the deuteron structure at finite $Q^2$ and at $x>1$.

\end{document}